\documentclass[twocolumn,preprintnumbers,amsmath,amssymb]{revtex4}
\usepackage{amssymb}
\usepackage{txfonts}

\usepackage{graphicx}
\usepackage{bm}

\begin{document}

\title{A wave-pulse neural network for quasi-quantum coding}

\author{Wen-Zhuo Zhang}
\email{sturmanz@hotmail.com}
\affiliation{AGI Lab, Qusute Ltd., Beijing, P. R. China}
\date{\today}

\begin{abstract}
We design a physical wave-pulse neural network (WPNN) for both wave and pulse propagation, which gives more degrees of freedom for neural coding than spike neural networks (SNN). We define the rules and the information entropy of this kind of neural network, where the signal speed, arrival time, and the length of connections between neurons all become crucial parameters for signal coding. We call it quasi-quantum coding (QQC) since the combination of wave and pulse signals here behaves like a classical mimic of quantum wave-particle duality, and can be studied by borrowing some concepts form quantum mechanics. We present that the quasi-quantum coding can give efficient methods for both sound and image recognitions. We also discuss the possibility of the wave-pulse neural network and the quasi-quantum coding methods running on it in biological brains where both neural oscillations and action potentials are important to cognition.
\end{abstract}

\maketitle

\section{introduction}

Quantum mechanics is a core theory of modern physics which dominates the behavior of microscopic world such as atoms, molecules and elementary particles. Whether quantum mechanics play an important role in the brain cognition is a debate for decades. Penrose guessed that quantum mechanics may play an role \cite{penrose}, while Tegmark argued that the quantum decoherence time is too short for any quantum information process in brain \cite{tegmark}. Recently, M. Fisher gave a hypothesis that the nuclear spins of Posner molecules in neural system may work as long lifetime qubits and form long-range quantum entanglements \cite{fisher}. However, the way such quantum information processes affecting real neural functions is still unknown.

In neural science, the neurons and their networks exist in a macroscopic scale which is dominated by classical physics rather than quantum physics since all quantum effects in a neuron should vanish due to quantum decoherence as Tegmark argued \cite{tegmark}. On the other hand, recent studies show that some quantum entanglement effects can be emulated even in a pure classical system of superposed waves \cite{classical,optical}. Therefore, even if the brain neural networks can present some quantum effect, it is more likely some kind of emulation (or mimic) of quantum system by classical wave system rather than a true microscopic quantum system.

In this paper, we define a kind of physical neural network with both wave and pulse propagating between neurons by six rules, and define its information entropy. It is evoked by the simultaneous propagation of brain-waves and action potential pulses in the brain, which are both crucial to brain memory and cognition \cite{phase}. In our neural network, any input signal can be coded by waves and pulses together and form unique neural connection patterns. The physical parameters such as the signal speed, the arrival time \cite{arrival}, and the length of connections between neurons become essential here. We choose the word "quasi-quantum" to call this coding method since 1) it is a classical mimic of the wave-particle duality in quantum mechanics, with wave and pulse part corresponding to the wave and particle property in quantum mechanics, respectively, and 2) we can borrow the language from quantum mechanics to define a neuron with wave and single pulse passing at a time as a "ground state" and a neuron with double pulse meeting at a time as an "excited state". This mimic has a similar starting point to the de Broglie's original polit-wave theory \cite{politwave}, which treat the wave and particle property independently rather than unify them as quantum mechanics does.

\section{the wave-pulse neural network and its information entropy}

We define the wave-pulse neural network (WPNN) with six rules: 1) the neural network is made up with sub-networks, and there are neuron-to-neuron connections in every sub-network and between every two sub-networks; 2) each sub-network only allows unique frequency wave signals to propagate, while pulse signals can propagate through all connections; 3) every wave signal has a constant amplitude, and every pulse signal have a constant amplitude too, while the amplitude of a pulse signal is much larger than the amplitude of a wave signal; 4) the temporal width of a pulse signal is much narrower than any period of a wave signal; 5) a neuron is excited when two pulse signals from different connections arrive at it simultaneously; 6) all signals propagate at a same speed.

\begin{figure}
\includegraphics[width=80mm]{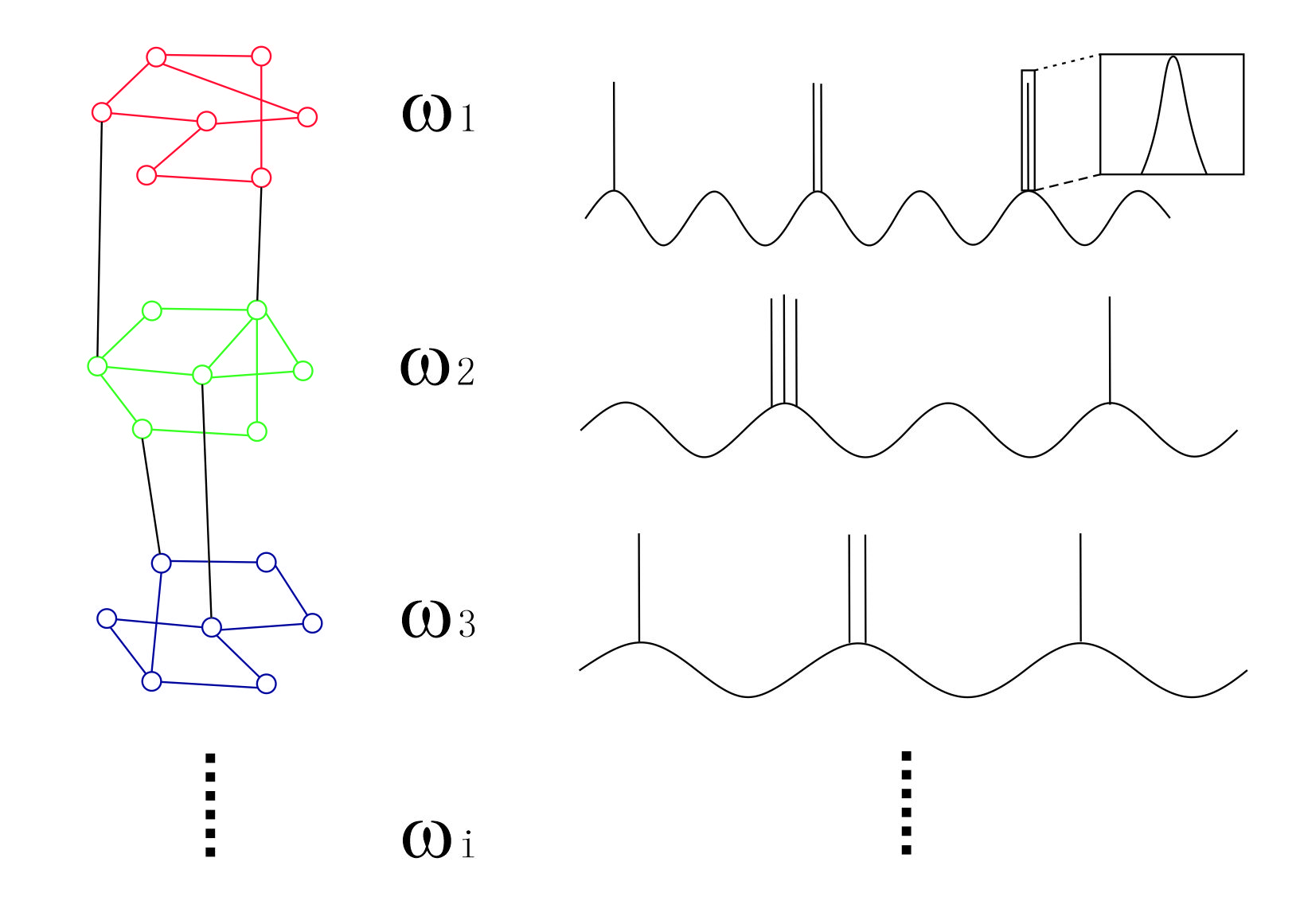}
\caption{The wave-pulse neural network is made up with sub-networks and each two sub-networks have neuron-to-neuron connection between them. The $\emph{i}$th sub-network only allow the wave signal with frequency $\omega_i$ to propagate, while all connections in the network allow pulse signals to propagate. Wave and pulse signals can overlap together. The duration of a pulse signal is much narrower than a period of a wave signal, while the amplitude of a pulse signal is much larger than it of a wave signal}\label{network}
\end{figure}

Fig. \ref{network} gives the schematic connections of the wave-pulse neural network. We call the $\emph{i}$th sub-network have a mode of $\omega_i$, which means that it only allow the wave signal with frequency $\omega_i$ to propagate. The $\emph{i}$th sub-network have $M_i$ neurons ($M_i>>1$), and the $\emph{j}$th neuron in it has $C_{ij}$ connections to other neurons within the sub-network.

For the $\emph{i}$th sub-network, we can define an information entropy
\begin{equation}\label{subentropy}
H_i=\frac{1}{M_i-1}\sum_{j=1}^{M_i}C_{ij}log_{M_i-1}C_{ij},
\end{equation}
where the information entropy equals to zero if $C_{ij}=0$ (no connections) or $C_{ij}=1$ (only paired connections). At the minimal network connection, each neuron connects two other neurons, then the information entropy becomes $H_i=\frac{1}{M_i-1}\sum_{j=1}^{M_i}2log_{M_i-1}2<<1$. At the maximum network connection, each neuron connects $M_i-1$ other neurons, then the information entropy is $H_i=1$.

We can generalize a sub-network's information entropy to the whole network. Assume that the network has $\emph{N}$ sub-networks and the $\emph{i}$th sub-network has $C_i$ connections to other sub-networks (neuron-to-neuron), then we can define the total information entropy
\begin{equation}\label{subentropy}
\begin{aligned}
H&=\frac{1}{N-1}\sum_{i=1}^{N}H_i{C_i}log_{N-1}{C_i}\\
&=\frac{1}{N-1}\sum_{i=1}^{N}({C_i}log_{N-1}{C_i})(\frac{1}{M_i-1}\sum_{j=1}^{M_i}C_{ij}log_{M_i-1}C_{ij}),
\end{aligned}
\end{equation}
where the information entropy equals to zero if $C_i=0$ (no connections between sub-networks) or $C_i=1$ (only paired connections between two sub-networks). At the minimal network connection, every sub-network connects two other sub-networks with only one neuron-to-neuron connection to each sub-network, and $0<H_i<<1$ for every sub-network, then the total information entropy is $H=\frac{1}{N-1}\sum_{i=1}^{N}H_i2log_{M_i-1}2<<1$. At the maximum network connection, each sub-network connects $N-1$ other sub-networks and $H_i=1$ for every sub-network, then the total information entropy of the network is $H=1$.

We see the information entropy above measures the uncertainty of wave and pulse signal's propagation. When $H=0$, there is no signal propagation (or only certain propagation between two paired neurons). At the minimal network connection, a signal in a neuron has two choices to propagate. At the maximum network connection, $H=1$, a signal in a neuron has maximum choices to propagate. Therefore, it can be considered as a generation of Shannon's information entropy that measures the uncertainty of bits.

\section{Quasi-quantum coding method on the network}

Since we define a wave-pulse neural network, a wave signal can propagate though all the connections in a sub-network while a pulse signal can propagate though all the connections in the whole network. Then which way is adopted for coding depends on whether it can excite a neuron. According to the rule 5 in section II, two pulse signals form different connections should arrive at a neuron simultaneously to excite it. We define it as an quasi-quantum excitation $|1_\omega>$ with a frequency $\omega$ from the local wave signal and a particle number $n=1$ from the pulse signals in order to mimic the wave-particle duality of a real quantum excitation. For an analog signal, we can code its frequencies by wave signals and code its amplitudes by pulse signals, and form a quasi-quantum signal.

For example, the $\emph{i}$th frequency component of an analog signal can be code into a quasi-quantum signal as
\begin{equation}\label{single}
M_i=A_w\cos(k_ix+\omega_it)+\sum_{n=1}^{N}A_pp_n(t_n),
\end{equation}
where $A_w$ is the constant amplitude of a wave signal and $A_p$ is the constant amplitude of a pulse signal as defined by rule 3 in section II. $k_i$ is the wavenumber and $\omega_i$ is the frequency of the wave signal in $\emph{i}$th sub-network which are corresponding to the frequency of the signal's $\emph{i}$th frequency component. $t_n$ is a series of discrete moments where the signal's $\emph{i}$th frequency component appears, and $p_n$ is the number of pulse signals at $t_n$ which is corresponding to the amplitude of the signal's $\emph{i}$th frequency component at $t_n$. Then any analog signal can be code into a series of quasi-quantum signals with decomposing its frequency components.

Here a quasi-quantum signal acts like a vacuum state $|0_\omega>$ in its sub-network, and every two pulse signals meeting each other act like a particle creation operator $a^\dagger$. When two pulse signals meet in a neuron, a quasi-quantum excitation $a^\dagger|0_\omega>=|1_\omega>$ appears and recorded, and the connections between any two neurons with quasi-quantum excitations $|1_\omega>$ are also recorded. Then an analog signal can be record by a unique connection pattern of neurons with quasi-quantum excitations $|1_\omega>$ over all frequency components it has.

Technically, all the quasi-quantum excitations $|1_\omega>$ and their time order can be recorded by a time-vs-frequency matrix which also imply the connection relations between the neurons with quasi-quantum excitations. Such wave-pulse neural network for quasi-quantum coding is different to any artificial neural networks in current computer science due to that the arrival time of pulses depends on the connection lengths between neurons and the speed of quasi-quantum signals, which are both physical parameters rather than virtual connections in computer science. In next two sections, we give two examples of the quasi-quantum coding method for both sounds and images, respectively.

\section{quasi-quantum coding for sounds}

Sound wave is a mechanic wave signal with frequencies and amplitudes distributing over time. In order to code a sound wave, we can decompose it into its frequency components and code them with wave signals in different mode sub-network respectively, while we code the amplitude of each frequency component with the number of separated pulse signals at the moment when the frequency component appears.

\begin{figure}
\includegraphics[width=80mm]{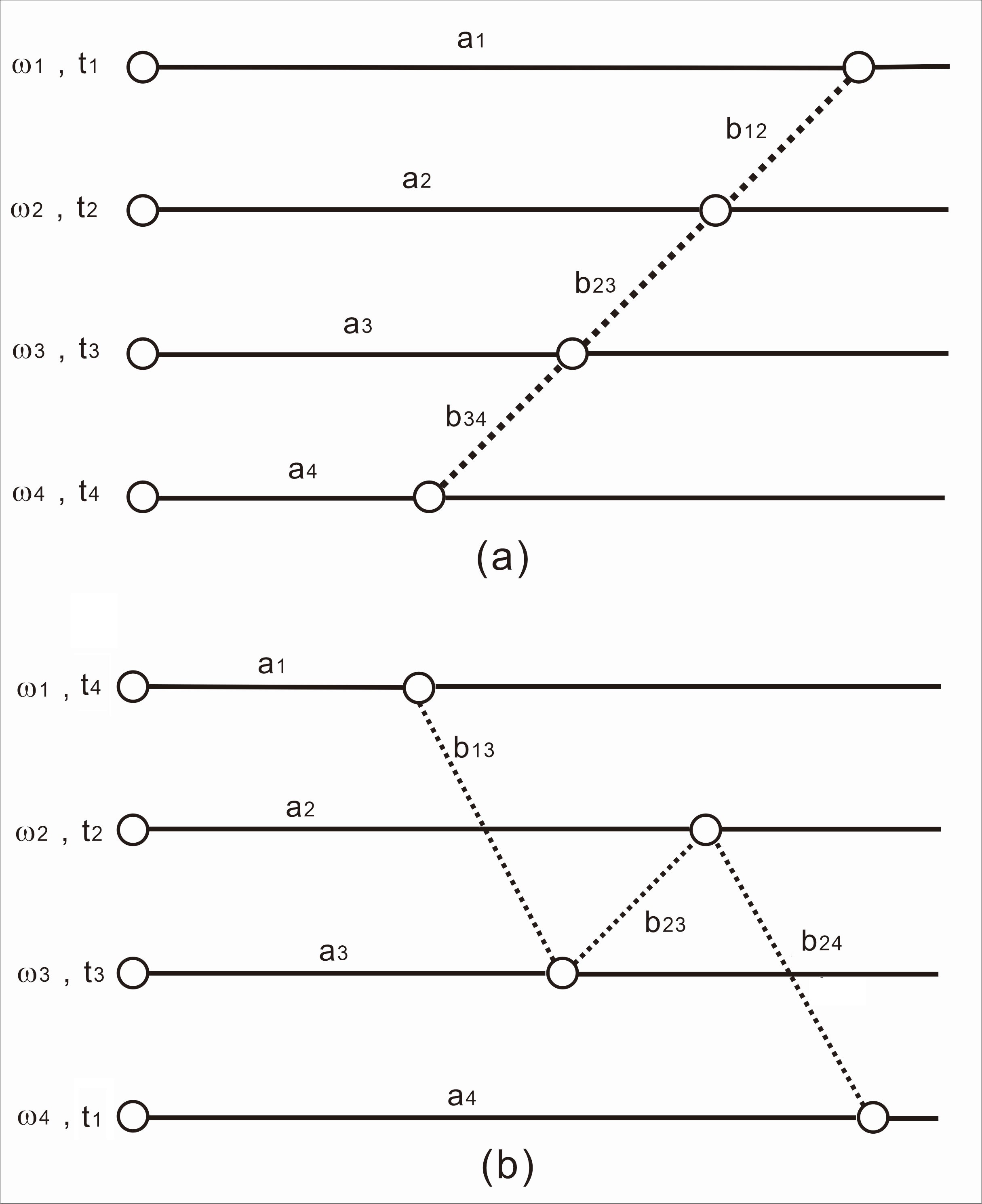}
\caption{Quasi-quantum coding method for simple sound signals. Solid lines represent to sub-networks, and dot lines represent to the connections between sub-networks for coding. (a) Connection pattern of neurons for coding a sound signal with four frequency components ($\omega_1$, $\omega_2$, $\omega_3$, $\omega_4$) appear in an entering time sequence ($t_1$, $t_2$, $t_3$, $t_4$); (b) Connection pattern of neurons for coding a sound signal with for frequency components ($\omega_4$, $\omega_2$, $\omega_3$, $\omega_1$) appear in an entering time sequence ($t_1$, $t_2$, $t_3$, $t_4$).} \label{soundcoding}
\end{figure}

Since a wave signal can travel only in its sub-network but pulse signals can travel along all connections, as rule 2 in section II, the connections between sub-network only have pulse signals. Besides, all signals travel at a same constant speed, as rule 6 in section II. Therefore, only the arrival time of pulse signals for the coding is necessary. Fig.\ref{soundcoding} gives a simple example of two pulses' meeting form two directions. According to rule 5 in section II, the neuron where the two pulse meet is a quasi-quantum excitation $|1_\omega>$. If the time interval of the two pulse signals is $\Delta_t$, the length difference between the two connection (start form the first neuron of each sub-network) is $d=v\Delta_t$, where $v$ is the constant speed of all signals.

For a simple sound signal with four frequency components where each frequency component appears at a different time, we can code it with four frequency ($\omega_1$, $\omega_2$, $\omega_3$, $\omega_4$) at a time order ($t_1$, $t_2$, $t_3$, $t_4$) that entering the network. In Fig.\ref{soundcoding}(a), a quasi-quantum signal travels in a sub-network $\omega_i$ with a distance $a_i$ and its pulse signals also travels from the sub-network $\omega_i$ to the sub-network $\omega_j$ with a distance $b_{ij}$. When the pulse form $\omega_1$ meets another pulse in $\omega_2$, the neuron where they meet is excited and the distances follows a relation $a_1+b_{12}-a_2=v(t_2-t_1)$. Next, when the pulse form $\omega_2$ meets another pulse in $\omega_3$, the relation is $a_2+b_{23}-a_3=v(t_3-t_2)$, and when the pulse form $\omega_3$ meets another pulse in $\omega_4$, the relation is $a_3+b_{34}-a_4=v(t_4-t_3)$. In another case, if we code a signal with four frequency ($\omega_4$, $\omega_2$, $\omega_3$, $\omega_1$) at a entering time sequence ($t_1$, $t_2$, $t_3$, $t_4$), the relations become $a_4+b_{24}-a_2=v(t_2-t_1)$, $a_2+b_{23}-a_3=v(t_3-t_2)$, and $a_3+b_{13}-a_1=v(t_3-t_1)$, as Fig.\ref{soundcoding}(b) shows.

This quasi-quantum coding method gives a unique network connection pattern of all excited neurons for a sound signal in an irreversible time order. When we record all the excited neurons (as well as the non-excited neuron in the first sub-network that connected to the first excited neuron), we can get a time-vs-frequency matrix, where the transverse direction is time order (from left to right) and the longitudinal direction are frequencies. For the simple example of Fig.\ref{soundcoding}, we can get two matrixes
\begin{equation}\label{four}
\left(
\begin{array}{cccc}
p_1 & 0 & 0 & 0 \\
0 & p_2 & 0 & 0 \\
0 & 0 & p_3& 0 \\
0 & 0 & 0 & p_4 \\
\end{array}
\right)\&
\left(
\begin{array}{cccc}
0 & 0 & 0 & p_1 \\
0 & p_2 & 0 & 0 \\
0 & 0 & p_3& 0 \\
p_4 & 0 & 0 & 0\\
\end{array}
\right).
\end{equation}
Here $p_i$ is the number of quasi-quantum excitations in the $\omega_i$ sub-network which is corresponding to the amplitude of the sound signal's frequency component (coded by $\omega_i$). If the two sound signals mix together, the mixed matrix from quasi-quantum coding would be
\begin{equation}\label{four}
\left(
\begin{array}{cccc}
p_1 & 0 & 0 & p_1 \\
0 & p_2+p_2 & 0 & 0 \\
0 & 0 & p_3+p_3& 0 \\
p_4 & 0 & 0 & p_4 \\
\end{array}
\right).
\end{equation}

All such matrixes have a translation invariance for both frequency and entering time of quasi-quantum signals. It means we can get a matrix for a sound signal ($\omega_1+\omega_0$, $\omega_2+\omega_0$, $\omega_3+\omega_0$, $\omega_4+\omega_0$, $\cdot\cdot\cdot$) at time order ($t_1+t_0$, $t_2+t_0$, $t_3+t_0$, $t_4+t_0$, $\cdot\cdot\cdot$) which is same to the matrix for the sound signal ($\omega_1$, $\omega_2$, $\omega_3$, $\omega_4$, $\cdot\cdot\cdot$) at time order ($t_1$, $t_2$, $t_3$, $t_4$, $\cdot\cdot\cdot$). The offset frequency "$\omega_0$" and offset time "$t_0$" are free to choose. So it makes the quasi-quantum coding method has symmetries of frequency translation, which is useful for cognising sounds with same syllables but different frequencies (pitch).

\section{quasi-quantum coding for images}

The quasi-quantum coding method can be applied to code images in a similar way to the coding of sounds. Since an image is a two dimensional spatial distribution of data, in order to code it into quasi-quantum signals on a wave-pulse neural network, we need to set a protocol. In digital computers, most of the images are coded line by line in order to transform two-dimensional spatial data into one-dimensional data over time. However, in our quasi-quantum coding method, we can not distribute every pixel a frequency since the pixel number of a image is usually large.

Inspired by the active scan ability of animal eyes (saccade) \cite{saccade}, we can use four individual wave-pulse neural networks to acquire signals form scanning. We define $NL$ as the network that receives and codes the signals form the left-to-right scan "$\rightarrow$", $NR$ as the network that receives and codes the signals form the right-to-left scan "$\leftarrow$", $NT$ as the network that receives and codes the signals form the top-to-bottom scan "$\downarrow$", and $NB$ as the network that receives and codes the signals form the bottom-to-top scan "$\uparrow$". Every network has sub-networks with different frequencies, and each frequency is corresponding to a unique row of pixels in $NL$ and $NR$, or a unique column of pixels in $NT$ and $NB$.

With this protocol, we can code any symbol by the quasi-quantum coding method on the four wave-pulse neural networks, and each network works in a same way to the coding of sounds where a frequency-vs-time matrix is finally recorded. For example, we can code a "T" type symbol into quasi-quantum signals and record four frequency-vs-time matrixes as
\begin{equation}\label{T-symbol}
\left(
\begin{array}{cccc}
p & \ldots & p \\
0 & \ldots & 0 \\
\vdots &  & \vdots \\
0 & \ldots & 0 \\
\end{array}
\right)\&
\left(
\begin{array}{cccc}
p & \ldots & p \\
0 & \ldots & 0 \\
\vdots &  & \vdots \\
0 & \ldots & 0 \\
\end{array}
\right)\&
\left(
\begin{array}{cccc}
0 & \ldots & 0 \\
\vdots &  & \vdots \\
0 & \ldots & 0 \\
p & \ldots & p \\
0 & \ldots & 0 \\
\vdots &  & \vdots \\
0 & \ldots & 0 \\
\end{array}
\right)\&
\left(
\begin{array}{cccc}
0 & \ldots & 0 \\
\vdots &  & \vdots \\
0 & \ldots & 0 \\
p & \ldots & p \\
0 & \ldots & 0 \\
\vdots &  & \vdots \\
0 & \ldots & 0 \\
\end{array}
\right).
\end{equation}
where the matrixes lie in a $NL \& NR \& NT \& NB$ way. $p$ is the number of pulses on every neuron with quasi-quantum excitations.

For a "+" type symbol, we can record four same frequency-vs-time matrixes as
\begin{equation}\label{plus}
\left(
\begin{array}{cccc}
0 & \ldots & 0 \\
\vdots &  & \vdots \\
0 & \ldots & 0 \\
p & \ldots & p \\
0 & \ldots & 0 \\
\vdots &  & \vdots \\
0 & \ldots & 0 \\
\end{array}
\right)\&
\left(
\begin{array}{cccc}
0 & \ldots & 0 \\
\vdots &  & \vdots \\
0 & \ldots & 0 \\
p & \ldots & p \\
0 & \ldots & 0 \\
\vdots &  & \vdots \\
0 & \ldots & 0 \\
\end{array}
\right)\&
\left(
\begin{array}{cccc}
0 & \ldots & 0 \\
\vdots &  & \vdots \\
0 & \ldots & 0 \\
p & \ldots & p \\
0 & \ldots & 0 \\
\vdots &  & \vdots \\
0 & \ldots & 0 \\
\end{array}
\right)\&
\left(
\begin{array}{cccc}
0 & \ldots & 0 \\
\vdots &  & \vdots \\
0 & \ldots & 0 \\
p & \ldots & p \\
0 & \ldots & 0 \\
\vdots &  & \vdots \\
0 & \ldots & 0 \\
\end{array}
\right),
\end{equation}
For a "X" type symbol, we can record four same frequency-vs-time matrixes as
\begin{equation}\label{X-symbol}
\begin{aligned}
\left(
\begin{array}{ccccc}
p & 0 & \ldots & 0 & p \\
0 & p & \vdots & p & 0 \\
\vdots & \vdots & \ddots & \vdots & \vdots \\
0 & p & \vdots & p & 0\\
p & 0 & \ldots & 0 & p \\
\end{array}
\right)\&
\left(
\begin{array}{ccccc}
p & 0 & \ldots & 0 & p \\
0 & p & \vdots & p & 0 \\
\vdots & \vdots & \ddots & \vdots & \vdots \\
0 & p & \vdots & p & 0\\
p & 0 & \ldots & 0 & p \\
\end{array}
\right)
\\
\&\left(
\begin{array}{ccccc}
p & 0 & \ldots & 0 & p \\
0 & p & \vdots & p & 0 \\
\vdots & \vdots & \ddots & \vdots & \vdots \\
0 & p & \vdots & p & 0\\
p & 0 & \ldots & 0 & p \\
\end{array}
\right)\&
\left(
\begin{array}{ccccc}
p & 0 & \ldots & 0 & p \\
0 & p & \vdots & p & 0 \\
\vdots & \vdots & \ddots & \vdots & \vdots \\
0 & p & \vdots & p & 0\\
p & 0 & \ldots & 0 & p \\
\end{array}
\right).
\end{aligned}
\end{equation}

For any curve, we can record four frequency-vs-time matrixes with $p$ distributes as a curve in each matrix. For example, we can get four same frequency-vs-time matrixes for an "O" type symbol, and in each matrix, the nonzero elements (pulse number) distribute like a circle (or a ellipse, depends on the spatial and temporal resolutions).

In our protocol, an image with $1000\times1000$ pixels require $4000$ sub-networks to code rather than to give each pixel a frequency that require $100,000$ sub-networks for this image. In order to make this coding method work, all the pixels of an image should be acquired parallel, which mean that we need to set a CCD or a CMOS camera to a parallel mode and lead signals of all pixels parallel into our physical networks.

For scale-invariant image cognition, our quasi-quantum coding method can use a camera with auto-focus mode, which can make similar images with different sizes to have similar inputs of pixels and finally get similar matrixes. We can also select a range of interest of each image to include the similar parts over all images, unify the pixels of range of interests (usually reduce the pixels proportionally to the lowest ones), and use quasi-quantum coding method to output similar matrixes.

\section{conclusion and discussions}

We design a wave-pulse neural network with six rules and information entropy. The information entropy is proportional to the connection complexity of the neural network, thus less connections mean less information entropy. As we know in neural science, a brain has more synapse connections between neurons in baby phase, and learning is a process of reducing connections between neurons \cite{brain}. So if the animal brain is (or partly is) a biological wave-pulse neural network, the learning process in a brain is a process of reducing information entropy.

We also define the quasi-quantum coding methods for sounds and images that work on the wave-pulse neural network. For coding sounds, we use a simplified quasi-quantum coding method where each frequency have a unique sub-network to propagate. A typical hearing range of a human is up to 20kHz. If we set the frequency resolution as 1Hz, a common sound signal may requires thousands of sub-networks to code. In order to code sound signals for a much wider frequency range with less sub-networks, we can set a binary activation of N sub-networks. For example, sixteen sub-networks have $2^{16}-1=65,535$ binary activation modes (with 1 means active and 0 means non-active for a sub-network), which can code sound signals up to 60kHz with 1Hz resolution.

For coding images, we use a quasi-quantum coding method with scanning a image for four time ("$\rightarrow$", "$\leftarrow$", "$\downarrow$" and "$\uparrow$"), and every scan is similar to the coding method sounds where each row or column of pixels is corresponding to a unique sub-network. When human eyes receive images, they make saccade that scanning the features of an image rapidly to get information \cite{saccade}. The scan traces of the saccade may oblique or even curved. In our quasi-quantum coding method, we have four rectangularly scan ways. Technically, we can superimpose any two of the four scan ways and change the scan speed of them individually to get any trace we want.

Since a signal can be coded into a unique connection pattern of neurons with quasi-quantum excitations $|1_\omega>$ in a wave-pulse neural network, it is satisfied with the memory mechanics of a biological neural network where a signal is stored by strengthening a unique connection among some neurons \cite{brain}. Besides, the quasi-quantum coding method for both sound and any scan of image are not time reversal. It is also satisfied with the fact that any memory in an animal brain is time-ordered \cite{brain}.

Therefore, it is interesting to test whether the wave-pulse neural network and the quasi-quantum coding methods running on it are involved in animal brain. In another way, artificial wave-pulse neural network could be built physically, or simulated on computers at first. More quasi-quantum coding methods besides coding sounds and images could be developed and run on these wave-pulse neural networks in order to benefit the development of artificial general intelligence.


\begin{thebibliography}{10}

\bibitem{penrose}
R. Penrose, The Emperor's New Mind, Oxford Univ. Press (1989); R. Penrose, in The Large, the Small and the Human Mind, Cambridge Univ. Press (1997).

\bibitem{tegmark}
M. Tegmark, Phys. Rev. E \textbf{61}, 4194 (2000).

\bibitem{fisher}
M. P. A. Fisher, arXiv:1508.05929.

\bibitem{classical}
B. R. La Cour and G. E. Ott, New J. Phys. \textbf{17} 053017 (2015).

\bibitem{optical}
X. -F. Qian, B. Little, J. C. Howell, and J. H. Eberly, Optica \textbf{2}, 611 (2015).

\bibitem{phase}
J. Fell and N. Axmacher, Nat. Rev. Neurosci. \textbf{12}, 105 (2011).

\bibitem{arrival}
S. J. Thorpe, Parallel processing in neural systems and computers 91-94 (1990).

\bibitem{politwave}
L. de Broglie, Found. Phys. \textbf{1}, 5 (1970).

\bibitem{saccade}
K. R. Gegenfurtner, Perception \textbf{45}, 1333 (2016).

\bibitem{brain}
M. F. Bear, B. W. Connors, M. A. Paradiso, Neuroscience: Exploring the Brain (4th ed), Jones \& Bartlett Learning (2015).

\end{thebibliography}
\end{document}